\documentstyle[psfig]{article}
\begin{document}

\title{Quantum Trajectories and Quantum Measurement Theory}  
\author{H. M. Wiseman  \\ Department of Physics, University of Auckland,
 Auckland, New Zealand \footnote{Current (2003) E-mail: H.Wiseman@griffith.edu.au}}
\date{published in Quantum Semiclass. Opt. {\bf 8}, 205-222 (1996)}
\maketitle

\newcommand{\beq}{\begin{equation}}
\newcommand{\eeq}{\end{equation}}
\newcommand{\bqa}{\begin{eqnarray}}
\newcommand{\eqa}{\end{eqnarray}}
\newcommand{\nn}{\nonumber}
\newcommand{\dg}{^\dagger}
\newcommand{\smallfrac}[2]{\mbox{$\frac{#1}{#2}$}}
\newcommand{\ket}[1]{| {#1} \rangle}
\newcommand{\bra}[1]{\langle {#1} |}
\newcommand{\sch}{Schr\"odinger }
\newcommand{\schs}{Schr\"odinger's }
\newcommand{\hei}{Heisenberg }
\newcommand{\heis}{Heisenberg's }
\newcommand{\half}{\smallfrac{1}{2}}
\newcommand{\bl}{{\bigl(}}
\newcommand{\br}{{\bigr)}}
\newcommand{\ito}{It\^o }
\newcommand{\str}{Stratonovich }
\newcommand{\bfi}{{\bf I}_{[0,t)}}

\begin{abstract}
Beyond their use as numerical tools, quantum trajectories can be ascribed a
degree of reality in terms of quantum measurement theory. In fact, they arise
naturally from considering continuous observation of a damped quantum system. A
particularly useful form of quantum trajectories is as linear (but non-unitary)
stochastic \sch equations. In the limit where a strong local oscillator is used
in the detection, and where the system is not driven, these quantum trajectories
can be solved. This gives an alternate derivation of the probability
distributions for completed homodyne and heterodyne detection schemes. It also
allows the previously intractable problem of real-time adaptive measurements to
be treated. The results for an analytically soluble example of adaptive phase
measurements are presented, and future developments discussed.

\end{abstract}

PACS: 42.50.Lc, 03.65.Bz, 42.50.Dv


\section{Introduction}

The concept of stochastic \sch equations ({\sc sse}s) \cite{DalCasMol92} or quantum
trajectories \cite{Car93b} has generated an impressive degree of activity in the
quantum optics community over the past few years \footnote{Related ideas have also
been developed by workers outside quantum optics. Space limitations prevent me from
discussing these.}. To one outside the field, it may be hard to synthesize all of
this activity, not least because there are (in my mind) three distinct
interpretations of quantum trajectories. I am here classifying authors by the
ontological status which they ascribe to their quantum trajectories: not real;
subjectively real; or objectively real. Alternatively, these three interpretations
could be stated as quantum trajectories being respectively: a mathematical tool; an
application of quantum measurement theory; or a theory of microscopic state
reduction. In the bulk of this paper I am concerned with quantum trajectories as a
part of quantum measurement theory (that is, with what I am calling the subjective
interpretation of quantum trajectories). Note that by ``measurement theory'' I mean
the body of mathematical techniques used to describe the process of measurement
within the standard Copenhagen interpretation \cite{Boh58,Hei59} of quantum
mechanics. I am not concerned with  metaphysical arguments regarding the adequacy of
the Copenhagen viewpoint, nor with physical extensions to standard quantum theory.
However, interpretational issues have been raised in the context of quantum
trajectories, and the potential confusion regarding the reality of {\sc sse}s may
obscure the results I wish to present. For these reasons, I will begin with a brief
review of the interpretations of quantum trajectories. I am not aware of a
comparison of this sort being in print.

\subsection{Non-Real Quantum Trajectories}

The ``non-real'' interpretation of quantum trajectories treats the 
trajectories simply as numerical tools for solving the master equation.
This is essentially the attitude taken by Dalibard, Castin and M\o lmer 
\cite{DalCasMol92} who introduced the concept, although they do mention that their 
``Monte Carlo Wavefunction''  approach also provides ``new physical 
pictures''. Note the emphasis on wavefunctions in their terminology.
The value of quantum trajectories for pure states as a computational tool 
is that it only takes $2N-2$ real numbers to store a state vector in an 
$N-$dimensional Hilbert space, compared to $N^2-1$ real numbers for a state 
matrix. Of course, this gain is offset by the fact that many trajectories 
are required to obtain a reliable ensemble average. However, for a large 
Hilbert space, quantum trajectories may offer a real advantage for 
numerical computation. One very practical application is computing 
the centre-of-mass motion of a fluorescent atom
\cite{MolCasDal93,DumZolRit92,MarDumTaiZol93,Mar93,HolMarMarZol95}. 

\subsection{Subjectively Real Quantum Trajectories}

In the body of this paper, quantum trajectories are attributed some degree of 
reality, in that they accurately model the evolution of a system conditioned on 
continuous observation. This can be called ``subjective reality'', 
because the nature of the trajectories depend on the measurement scheme 
chosen by the observer. Once a particular measurement scheme has been
chosen, the conditioned evolution of the system (as modeled by quantum trajectories)
is objectively real in the sense that all observers will agree on the results. The
subjectivity is present in the sense that it is not permissible to make a
statement such as ``an excited atom in free space will spontaneously emit a photon
at a random point in time'' unless there is a photodetector to record the emitted
photon. A different detection scheme, such as heterodyne detection, will register
a continuous flow of radiation from the atom, not a jump. In other words, the
stochastic system evolution is only real in so far as the system is a subject of
observation; it is not real for the system as an object in an observerless universe.
As well as trajectories arising from different types of measurements, trajectories
which do not preserve pure  states (because of imperfect detection or bath
preparation) also arise in this interpretation.  

It is the subjective interpretation 
of quantum trajectories, in addition to their computational convenience,  which is the
motivation for the work by Carmichael and co-workers 
\cite{Car93b,AlsCar91,TiaCar92,Car93a,CarKocTia94,KocCarMorRai95}. Most
of this work was numerical, but analytic results were obtained for  some particular
systems and particular measurement schemes  \cite{Car93a,CarKocTia94}. Both quantum 
jump (direct detection) and quantum diffusion (homodyne detection) trajectories were
used. The measurement theory interpretation was also pointed out by Gardiner,
Parkins, and Zoller \cite{GarParZol92} for the case of direct detection (which was
formulated much earlier by Srinivas and Davies \cite{SriDav81}). 

There are a number of reasons for wishing to consider the measurement 
interpretation of quantum trajectories. Firstly, the simulated photocurrents 
can be used to calculate such quantities as the waiting time distribution for 
photodetections, which are quite difficult to calculate by traditional 
methods \cite{CarSinVyaRic89}. Another example is the all-or-nothing fluorescence
exhibited in the electron shelving of a single atom \cite{Coo88,GagMil93}. The  
technique of linear quantum trajectories recently published by Goetsch and Graham
\cite{GoeGra94b} has opened the possibility for at least partly analytical
solutions of measurement theory problems using quantum trajectories. This is the
subject of the body of this paper, and I will thus defer further discussion until
there. 

A second motivation is  that subjectively real
quantum trajectories give valuable  physical insight into irreversible quantum
processes, different from that offered by the master equation. For example, a
system as simple as a stationary two-level atom shows strikingly different quantum
trajectory evolution depending on the chosen method of detection, and that this
evolution sheds light on the respective  measurement results \cite{WisMil93c}.
Another example is that coherences which are lost in the non-selective (master 
equation) picture are seen to be merely distributed among different  quantum
trajectories \cite{Car93b,AlsCar91,CarKocTia94,GoeGraHaa95a,GarKni94}. These
reconstructed coherences can restore noisy oscillatory dynamics to a
system which would otherwise exhibit only damped oscillations
\cite{QuaColWal95}.

Thirdly, the measured  results can be used to select certain individual systems from
the  ensemble. In this way, it may be possible to see the coherences which are  lost
in the non-selective ensemble evolution  \cite{CarKocTia94,KocCarMorRai95}.
This could be looked upon as a very  basic sort of feedback, in which the measured
result is used to eliminate  some members of the ensemble. The use of quantum
trajectories in more active feedback schemes is a fourth motivation for subjectively
real quantum trajectories, which has been pursued by myself and Milburn and others 
\cite{WisMil93b,WisMil94a,Wis94a,WisMil94b,LieMil95,Wis95a,Wis95b}.

\subsection{Objectively Real Quantum Trajectories}

The third attitude towards quantum trajectories is a belief in their 
objective existence, independent of any measurement scheme. This belief 
is presumably founded on mistrust of the standard interpretations of 
quantum mechanics, and a desire to replace subjective collapses due to 
measurements with objectively real collapses. The roots of this approach 
are the dynamical state reduction models of Gisin \cite{Gis89}, 
Diosi \cite{Dio88}, Pearle \cite{Pea89} and Ghirardi-Rimini-Weber \cite{GRW},
which in turn hark back to the original objective interpretation of the 
wavefunction by \sch \cite{Roh87}. In this desire, proponents of objectively 
real quantum trajectory theory are akin to hidden variable theorists, but 
unlike the latter, they accept the wavefunction as a 
representation of reality. It is only when irreversible processes occur 
(described by a master equation) that stochasticity enters quantum 
mechanics, giving rise to quantum trajectories. Because the state vector is considered
a basic element of reality in this  interpretation, it only makes sense to consider
quantum trajectories  which preserve pure states. Thus it is that stochastic \sch
equations (as opposed to the more general stochastic master equations of subjective
quantum trajectories) are essential only if one regards them as not real or
objectively  real.

At least three pairs of authors have published on this theme. The first is Teich and
Mahler (TM). They published a single paper \cite{TeiMah92} presenting a theory of
quantum jumps in fluorescent atoms. It has been shown to be nonphysical by
myself and Milburn in reference \cite{WisMil93c}. The second pair, Gisin  and
Percival (GP), have published many papers \cite{GisPer92a,GisPer92b,GisPer93}, also
with other authors \cite{Gis93}, and have received considerable attention.
As pointed out by myself and Milburn,
the quantum trajectory which GP consider can be given an interpretation
in terms of unit-efficiency heterodyne detection (or ambi-quadrature homodyne
detection) of the electromagnetic field radiated by the system \cite{WisMil93c}.
However GP  attribute an objective reality to their {\sc sse}, saying that
it ``represents the evolution 
of an individual quantum system in interaction with its environment''. 
By adding {\em ad hoc} irreversible processes to a model of a three 
level atom, they are able to induce their equation to exhibit  
jump-like behaviour \cite{Gis93}, as seen in electron shelving 
experiments \cite{Coo88}. In general, 
however, GP cannot associate their trajectories with the detection of  
photons for the very good reason that they actually correspond to heterodyne, 
not direct detection. A third pair, Breuer and Petruccione (BP), have very recently 
published two papers  \cite{BrePet95a,BrePet95b} in which they claim to show that the
quantum jump  unraveling of reference \cite{GarParZol92} is {\em the} {\sc sse} for
a Markovian open quantum system. Although it is in many ways the most natural 
unraveling, the quantum jump {\sc sse} (which under some circumstances corresponds
to direct detection \cite{GarParZol92,WisMil93c}) does not have any fundamental
privileged status. BP derive it by choosing a particular basis in which to
diagonalize the bath, namely its energy eigenstates.

The basic problem with the objective interpretation of quantum 
trajectories lies with the origin of irreversible evolution. In standard
quantum optics, the irreversibility of the master equation for a system 
is an approximation to the exact reversible evolution for the system and its
environment \cite{Gar91}. None of the above authors try to define in what contexts
the master equation  evolution is meant to be truly irreversible. If there were some
fundamental physical mechanism by  which irreversibility entered the world, for
example at some sufficiently  large scale, then it is conceivable that one of the
three schemes outlined above could be correct \footnote{Of course, these are not the
only three conceivable dynamical  state reduction schemes; two which have gained
considerable attention are those of GRW \cite{GRW} and Penrose \cite{Pen90}.}. The
quantum measurement ``problem'' would then be solved,  as measurement results would
be determined by the irreducibly stochastic quantum  evolution. Irrespective of this
speculation, the models of TM, GP and BP cannot be  objectively real for the systems
to which the authors apply them \footnote{These criticism now appears to be accepted
by Percival, as evidenced by reference \cite{Per94}.}. Irreversible stochastic
evolution, in the manner of TM, GP or BP, does not take place within an individual
atom. The entanglement between an atom and its outgoing field  is a fact which is not
represented by the models of TM, GP and BP. This  entanglement is most simply shown
by the fact that one can choose different detection schemes to observe the atom, as
discussed in \cite{WisMil93c}. Gisin and Percival's scheme only gives the correct
results for heterodyne detection, Breuer and Petruccione's for direct detection,
while Teich and Mahler's scheme will not give the correct results for any detection
scheme \cite{WisMil93c}.

\subsection{Outline}

Having discussed the non-real and objectively real interpretations of quantum
trajectories, the remainder of this paper is devoted to quantum trajectories as
a part of quantum measurement theory. I begin with a discussion of traditional
quantum measurement theory, and why this is inadequate to describe practical
measurements. This leads to an outline of the theory of operations and effects,
and the different ways in which probability can be incorporated in the theory.
Sec.~3 introduces quantum trajectories as an application of quantum measurement
theory to continuous observation. The description of such trajectories as linear
stochastic \sch equations is covered in Sec.~4. Sec.~5 presents the general
solution for the linear {\sc sse} for the case of an undriven system with a large local
oscillator used in the detection process. This allows a new derivation of the
operations and effects for completed homodyne and heterodyne measurements. In
Sec.~6 I apply the results of Sec.~5 to a measurement scheme which has not
previously been considered and which is probably intractable without linear
quantum trajectories. The measurements which I am describing here are adaptive
measurements, in which the past photocurrent is used to modify the present
conditions of the measurement. 
Here I discuss the general approach, and present results for the simple but
non-trivial example of adaptive phase measurements of a field containing at most
one photon. Sec.~7 concludes.

\section{Quantum Measurement Theory}

Traditional quantum measurement theory is founded on projective measurements.
Consider a measurement of fixed duration $T$ with a set of $N$ discrete possible
measurement results labeled by $r$. With each result $r$ there is an
associated projection operator $\Pi_r(T)$, such that
$\Pi_r(T)\Pi_s(T)=\Pi_r(T)\delta_{r,s}$. If the initial state of the system
(assumed pure for simplicity) is $\ket{\psi(t)}$, then the probability for
obtaining the result $r$ is \beq
P_r(T) = \bra{\psi(t)}\Pi_r(T)\ket{\psi(t)}.
\eeq
This can also be written
\beq
P_r(T) = \bra{\tilde{\psi}_r(t+T)}\tilde{\psi}_r(t+T)\rangle,
\eeq
where the tilde denotes an unnormalized state vector
\beq
\ket{\tilde{\psi}_r(t+T)} = \Pi_r(T)\ket{\psi(t)}.
\eeq
Furthermore, the final state of the system is simply this state, normalized if
desired 
\beq
\ket{\psi_r(t+T)} = \frac{\ket{\tilde{\psi}_r(t+T)}}{\sqrt{P_r(T)}}.
\eeq

Although this theory may be fundamentally correct in some sense, it is inadequate
to describe practical measurements such as those in quantum optics. That is
because the experimenter never makes a direct measurement on the system of
interest. Rather, the system of interest (such as an atom) interacts with its
environment (the continuum of electromagnetic field modes), and the experimenter
observes the radiated field. Of course one could argue that the
experimenter does not observe the radiated field, rather that the field interacts
with a photodetector, which triggers a current in a circuit, which is coupled to a
CRO, which radiates more photons, which interact with the experimenter's
retina, and so on up the von-Neumann chain \cite{Von32}. The point is that at some
stage one has to consider the measurement as being made. If one attempts to apply a
projection postulate directly to the atom, one will obtain nonsensical predictions.
However, it can be shown that assuming a projective measurement of the field will
yield results negligibly different from those assuming a projective measurement at
any later stage, because of the rapid decoherence of macroscopic
objects \cite{Zur82}. For this reason, it  is sufficient to consider the field as
being measured. Because the field has interacted with the system, their quantum
states are entangled. If the initial state of the field is assumed known (which is
not unreasonable in practice, because it is often close to the vacuum state), then
the projective measurement of the field results in a measurement of the atom. Such
a measurement however is not projective \footnote{This is why the original quantum
Zeno paradox of Misra and Sudarshan \cite{MisSud77} can be avoided in practice.}.
Instead, we need a more general formalism to describe such measurements. This is
called the theory of operations and effects  \cite{Dav76}, which I will now present
in brief.

Say the initial system state vector is $\ket{\psi(t)}$ as above, and that the
measurement results are again labeled by $r$. Each $r$ is again associated with an
operator $\Omega_r(T)$, but it need not be a projection operator. I will call these
{\em measurement operators}.  The {\em effect} \cite{Dav76} for the result $r$ is
defined by  \beq
F_r(T) = \Omega_r(T)\dg \Omega_r(T). \label{hereisfac}
\eeq
This operator is used to generate the probability for the result via
\beq
P_r(T) = \bra{\psi(t)}F_r(T)\ket{\psi(t)}. \label{onea}
\eeq
Conservation of probability yields the single restriction which exists for the
measurement operators \cite{Dav76}, namely
\beq \label{coef}
\sum_r F_r(T) = 1.
\eeq
The set of all effects $\{ F(r) \}$ constitutes a positive-operator-valued
measure ({\sc povm}) on the space of results $r$ \cite{Dav76}. The state of the system
given the result $r$ is \beq
\ket{\psi_r(t+T)} = \frac{\Omega_r(T) \ket{\psi(t)}}{\sqrt{P_r(T)}}.
\label{oneb} \eeq

If one were doing only a single measurement, then the conditioned state
$\ket{\psi_r}$ would be irrelevant. However one often wishes to consider
a sequence of measurements, in which case the conditioned system state is vital. In
terms of the state matrix $\rho$, which allows the possibility of mixed initial
states, the conditioned state is 
\beq \rho_r(t+T) = \frac{{\cal J}[\Omega_r(T)]\rho(t)}{P_r(T)}, \eeq 
where $P_r(T)={\rm Tr}[\rho(t)F_r(T)]$ and for an arbitrary operator $a$, 
\beq {\cal J}[a]\rho = a\rho a\dg. \label{defcalJ}
\eeq
The superoperator ${\cal J}[\Omega_r(T)]$ is known as the {\em operation} for $r$
\cite{Dav76}. If the measurement were performed but the result ignored, the final
state of the system would be
\beq \label{rop}
\rho(t+T) = \sum_r P_r(T) \rho_r(t+T).
\eeq

Consider a hypothetical experiment involving a sequence of
measurements for which one knows the set of measurement operators $\{\Omega_r(T)\}$.
As a theorist, one may be interested in knowing {\em a}) the probability for
obtaining the possible measurement sequences, and {\em b}) the state of the system
during and at the end of the measurement, conditioned on the results obtained. If
initially $\rho(t)=\ket{\psi(t)}\bra{\psi(t)}$ then the above formalism shows that
one could obtain this information as follows. Start with $\ket{\psi(t)}$, and generate
the probabilities for the results using equation (\ref{onea}). One would then choose
results at random using these probabilities, and see what the new conditioned state
of the system is using equation (\ref{oneb}). This is the input into the next
measurement. The probability for each sequence of results to actually occur is
simply the probability that it was generated by the simulation procedure. The
simulation is repeated as may times as necessary to obtain good statistics. Call
this method A.

As in the case for projective measurements, this theory can be alternatively
stated as follows. The final state of the system conditioned on the result $r$ is
represented by an unnormalized state vector
 \beq \label{twoa}
\ket{\tilde{\psi}_r(t+T)} = \Omega_r(T) \ket{\psi(t)}.
\eeq
The probability for this result having been obtained is  then
\beq \label{twob}
P_r(T) = \bra{\tilde{\psi}_r(t+T)}\tilde{\psi}_r(t+T)\rangle.
\eeq
The spirit of this formulation is that one could simulate the measurement process
by randomly generating a result $r$ with probability $1/N$, then calculating
the appropriate unnormalized conditioned state. This is the input into the next
measurement step. The actual probability assigned to each sequence of measurement
results is the final norm squared of the state vector (\ref{twob}), where here
 $T$ would be the total time for the sequence of measurements.
This simulation would be less efficient because all measurement results are given
equal computational time, even those with vanishingly small probabilities. However
it would have the advantage that probabilities need to be calculated only once (at
the end). Call this method B.

In fact, it is possible to formulate the theory in any number of ways which
include these two as special cases. Define an (in general unnormalized) conditioned
state vector by
\beq \label{threea}
\ket{\bar{\psi}_r(t+T)} = \frac{\Omega_r(T)\ket{\psi(t)}}{\sqrt{\Lambda_r(T)}} .
\eeq
Here the $\Lambda_r(T)$ are positive numbers chosen such that
\beq
\sum_r \Lambda_r(T) = 1.
\eeq
They can thus be interpreted as probabilities in some sense, and I will call
$\Lambda_r$ the {\em ostensible} probability for the result $r$. The
{\em actual} probability for getting result $r$ is 
\beq \label{threeb}
P_r(T) = \Lambda_r(T)\bra{\bar{\psi}_r(t+T)}\bar{\psi}_r(t+T)\rangle.
\eeq
If the $\Lambda_r(T)$ are chosen such that $\Lambda_r(T)$ happens to equal
$\bra{\psi(t)}F_r(T)\ket{\psi(t)}$, then this formulation is equivalent to the
method A given above, with $\ket{\psi_r(t+T)} = \ket{\bar{\psi}_r(t+T)}$. If they
are chosen such that $\Lambda_r(T) = 1/N$, then this formulation is equivalent to
the method B, with
$\ket{\tilde\psi_r(t+T)}=\ket{\bar\psi_r(t+T)}\sqrt{N}$.  In general, one could do
the simulation of the measurement 
by choosing the results $r$ with probability $\Lambda_r(T)$, then calculating the
conditioned state vector $\ket{\bar{\psi}_r(t+T)}$. Each result would then be
weighted by $\bra{\bar{\psi}_r(t+T)}\bar{\psi}_r(t+T)\rangle$, where again $T$ would
be the total time. Because of the probability of generation, this weighting gives the
correct overall weighting to each member of the ensemble as given in equation
(\ref{threeb}). Call this method C.

This flexibility of formulation has been known for some time (see Goetsch and
Graham \cite{GoeGra94b} and references therein). As presented here it must appear
rather pointless. However we are about to consider quantum trajectories, as an
application of measurement theory for continuously (in time) monitored systems.
In this context, it turns out that the ability to separate out the probability
of each possible result into two components (a probability of generation
$\Lambda_r(T)$ and a weighting $\bra{\bar{\psi}_r(t+T)}\bar{\psi}_r(t+T)\rangle$),
will be of enormous use in solving the quantum trajectories. 

Before
turning to the topic of quantum trajectories, there is one more useful fact which can
be explained in terms of the general formalism. Recall that the final state of the
system if all measurement results are ignored is given by equation (\ref{rop}). This
can also be written as \beq
\rho(t+T) = \sum_r {\cal J}[\Omega_r] \rho(t).
\eeq
Now if we define a unitary rearrangement of of the measurement operators
$\Omega_r$ by
\beq
\Omega'_r = \sum_s U_{r,s} \Omega_s,
\eeq
where $U$ is a $c-$number matrix satisfying
$
\sum_r U_{r,s} U^*_{r,q} = \delta_{s,q}.
$
Then the unconditioned final state under the new measurement operators
$\{\Omega'_r\}$ is
 \begin{eqnarray}
\sum_r \Omega'_{r} \rho(t) {\Omega'_r}\dg &=& 
\sum_{r,s,q} U_{r,s} \Omega_s \rho(t)
\Omega_{q}\dg U^*_{r,q} \\
&=& \sum_{s,q} \delta_{s,q} 
\Omega_s\rho(t)\Omega_{q}\dg \; = \; \sum_s 
\Omega_s\rho(t)\Omega_{s}\dg . \label{aeie}
\end{eqnarray}
which is the same as that under the old measurement operators
$\{\Omega_r\}$ . That is to  say, one can
unitarily rearrange the measurement operators without changing the non-selective
system evolution.

\section{Quantum trajectories}

One case of the above measurement theory which is of great 
importance is continuous observation, with an infinitesimal 
measurement interval $T=dt$. The form of such a measurement 
theory is restricted by the requirement that the non-selective evolution
generated by the measurements should give a valid evolution equation 
for the state of the system. Explicitly, this non-selective evolution is 
\beq \label{markovmeast}
\rho(t+dt) = \sum_{r} \Omega_{r}(dt) \rho(t) 
\Omega_{r}\dg (dt).
\eeq
This is obviously Markovian \cite{Gar85}, with the increment in the 
system state from time $t$ to time $t+dt$ depending only on the state 
of the system at time $t$. A Markovian 
evolution equation for the state matrix of a system is called a master 
equation. Physically, it can be derived from the interaction of the 
system with its environment, if the environment can be treated as a 
bath. That is to say, it is necessary for the environment to irreversibly 
carry away information from the system. Alternatively, one can view the
environment  as a continuous stream of independently prepared apparatuses 
which are wheeled up to the system, interact for an infinitesimal time 
$dt$, and are wheeled away again to be measured. Although this 
sounds unrealistic, the electric-dipole interaction between a system
and the electromagnetic vacuum behaves like this to a good
approximation \cite{GarParZol92,Wis95b}.

The most general form of master equation is the so-called Lindblad form
\cite{Lin76,Gar91}
\begin{equation} \label{241me1} \label{Lin}
\dot{\rho} = -{i}[H,\rho] + \sum_\mu  {\cal D}[c_\mu] \rho.
\end{equation}
Here $H$ is an Hermitian operator and ${\cal D}$ is a 
superoperator taking an operator argument (enclosed in square brackets). It is 
defined for an arbitrary operator $r$ by
\beq \label{defcalD}
{\cal D}[r]\rho \equiv r\rho r\dg - \smallfrac{1}{2} (r\dg r \rho + \rho r\dg r).
\eeq
The operators $c_\mu$ are completely arbitrary. 

Consider now a special case where there is but one Lindblad operator $c$ for the
system, so that 
\beq \label{241me2}
\dot{\rho} = -i[H,\rho] + {\cal D}[c]\rho \equiv {\cal L}\rho.
\eeq
 Then it is possible to write the master
equation as the nonselective evolution due to a measurement [as in equation
(\ref{markovmeast})] with just two measurement operators $\Omega_r(dt)$ by choosing
\begin{eqnarray} \Omega_1(dt) &=& \sqrt{dt}\, c , \\
\Omega_0(dt) &= &1 - \left( {i}H + \smallfrac{1}{2}c\dg c \right) dt.
\end{eqnarray}
This is because 
\begin{equation} \label{me2a}
\rho(t+dt) = \sum_{r=0,1} {\cal J}[\Omega_r(dt)] \rho(t) 
 = (1+{\cal L}dt)\rho(t),
\end{equation}
where ${\cal L}$ is as given in equation (\ref{241me2}). 

With the measurement operators $\Omega_0(dt),\Omega_1(dt)$, 
it is evident that the measurement record will be a point process \cite{CoxIsh80},
as I will now explain. For almost all infinitesimal time intervals, the  measurement
result is $r = 0$, because $P_0(dt) = 1 - O(dt)$. The result $r=0$ is thus
regarded as a  null result. In the case of no result, the system state changes
infinitesimally, but not unitarily, via the operator  $\Omega_0$. At randomly
determined (but not necessarily Poisson  distributed) times, there is a result $r =
1$, which I will call  a {\em detection}. When this occurs, the system undergoes a
finite  evolution induced by the operator $\Omega_1$. This change can validly be 
called a {\em quantum jump}, although it must be remembered that it  represents a
sudden change in the observer's knowledge, not an  objective physical event as in
Bohr's original conception  \cite{Boh13}. Real measurements 
which correspond approximately to this ideal measurement model are made routinely 
in experimental quantum 
optics. If $c$ is the lowering operator for the dipole of the quantum system,
multiplied by the  square root of the damping rate, then this theory describes 
the  system evolution in terms of photodetections \cite{Wis95b}. 

Now these measurement operators can be unitarily rearranged by the matrix
\beq \label{urm}
\left( 
\begin{array}{cc}
	U_{00} & U_{01}  \\
	U_{10} & U_{11}
\end{array} \right) = \left( 
\begin{array}{cc}
	1-\half |\gamma|^2 dt & \gamma\sqrt{dt}  \\
	-\gamma^*\sqrt{dt} & 1-\half |\gamma|^2 dt
\end{array} \right).
\eeq
This is equivalent to the transformation
\begin{equation} \label{241transf}
 c \to c+\gamma \; ; \;\; H \to H -{i} \smallfrac{1}{2} 
(\gamma^*c - \gamma c^\dagger),
\end{equation}
under which the master equation (\ref{241me2}) is of course invariant. This
transformation preserves the property that the collapse operator $\Omega_1$
is infinitesimal, and the smooth evolution operator $\Omega_0$ is close to unity.
Explicitly,
\begin{eqnarray}
	\Omega_1(dt) & = & \sqrt{dt}(c+\gamma),
	\label{421om1} \\
	\Omega_0(dt) & = & 1 - dt\left[{i}H + 
	\smallfrac{1}{2}(c\gamma^*-c\dg\gamma) + 
	\smallfrac{1}{2}(c\dg+\gamma^*)(c+\gamma) \right].
	\label{421om0}
\end{eqnarray}
Physically, it amounts to adding a coherent field to the output
radiation field before it is detected by the system. This can be done using a
low-reflectance beam-splitter. In the limit $|\gamma|^2 \gg \langle c\dg c
\rangle$, this procedure is known as homodyne detection.

The specification of the measurement operators is all that is necessary to
define the measurement. However, it is useful to give a more explicit
description of the evolution of the system under this scheme. This is done by
introducing a stochastic increment $dN(t)$ which equals one if the result 1 is
obtained in the time interval $(t,t+dt)$ and 0 otherwise. Formally, $dN(t)$ is
defined by
\begin{eqnarray}
dN(t)^2 &=& dN(t) ,\\
	{\rm E}[dN(t)]&=&\langle \Omega_1\dg(dt)\Omega_1(dt) \rangle = dt \bra{\psi(t)}
(c+\gamma)\dg (c+\gamma) \ket{\psi(t)},
\end{eqnarray}
where a classical expectation value is denoted by ${\rm E}$ and the quantum
expectation value by angle brackets. 
From the measurement operators (\ref{421om1},\ref{421om0}), 
the stochastic evolution equation for the conditioned state vector is
\begin{eqnarray} 
d|\psi (t) \rangle  &=& \left[ dN(t)\left( \frac{c+\gamma}
{\sqrt{\langle (c\dg+\gamma^*)(c+\gamma)\rangle(t)}} - 1 \right)\right. \nn \\
&& \left. + \; dt \left( \frac{\langle 
c^\dagger c \rangle(t)}{2} - \frac{c^\dagger c}{2} + 
\frac{\langle c\dg \gamma + \gamma^* c \rangle(t)}{2} 
- \gamma^* c -{i}H \right) \right] |\psi(t) \rangle. \label{412sse1b}
\end{eqnarray}
 This gives the quantum trajectory for the system explicitly as a non-linear
stochastic \sch equation. Putting $\gamma=0$ gives the form generally used for
numerical solutions of the master equation.

\section{Linear Quantum Trajectories}

The evolution described by equation (\ref{412sse1b}) corresponds to the method A of
Sec.~2. That is to say, the results $dN(t)$ are generated with the
distribution they would have in actuality. No other weighting is necessary. In this
case, method B would be outstandingly inefficient. It would imply either generating
or not generating a photodetection with equal probability for every instant of time
(${\rm E}[dN]=\half$). The norms of the vast majority of states would rapidly become
vanishingly small, and their contribution to the ensemble would be negligible. The
only states which would significantly contribute to the ensemble would be those in
which photodetections were sparse, as they would be in actuality. However, there is a
half-way house, corresponding to method C of Sec.~2. This would be achieved by
keeping ${\rm E}[dN]=O(dt)$. Specifically, a convenient choice is \beq
\Lambda_1(dt) = {\rm E}[dN] = |\gamma|^2 dt,
\eeq
so that by necessity $\Lambda_0(dt) = 1 - |\gamma|^2 dt$.
Then the unnormalized conditioned system state would evolve via
\bqa
\ket{\bar\psi_1(t+dt)} &=&
\frac{\Omega_1(dt)\ket{\bar\psi(t)}}{\sqrt{\Lambda_1(dt)}}=\left(1+\frac{c}{\gamma}
\right) \ket{\psi(t)} \\ 
\ket{\bar\psi_0(t+dt)} &=&
\frac{\Omega_0(dt)\ket{\bar\psi(t)}}{\sqrt{\Lambda_0(dt)}}=\left[1 - dt
\left(\half c\dg c+ c \gamma^*+ iH \right) \right]\ket{\psi(t)} 
\eqa
The ostensible statistics for $N(t)$ are those of a Poisson process. 
The actual probability for a detection varies in time according to
\beq
P_1(dt) = \Lambda_1(dt)  \bra{\bar\psi_1(t+dt)}
\bar\psi_1(t+dt)\rangle = dt \bra{\psi(t)}
(c+\gamma)\dg (c+\gamma) \ket{\psi(t)},
\eeq
as before.

Writing out this evolution explicitly, we have
\beq \label{linqt1}
d\ket{\bar\psi(t)} = \left[ dN(t) \frac{c}{\gamma} - dt \left( iH + \half c\dg c 
+ \gamma^* c \right) \right] \ket{\bar\psi(t)}.
\eeq
This is considerably simpler than equation (\ref{412sse1b}), and, more importantly,
it is linear in $\ket{\bar\psi(t)}$. This has been achieved by transferring some
of the information about the probability of each measurement record into the norm
of $\ket{\bar\psi(t)}$. The choice of
$|\gamma|^2 dt$ for $\Lambda_1(dt)$ is obviously not a good one if the local
oscillator amplitude $\gamma$ vanishes. In the remainder of this paper I am concerned
with the opposite limit, $|\gamma| \to \infty$. Consider a short time $\delta t \ll
1$ in which the system changes negligibly, but the number of detections $\delta N(t)
\sim |\gamma|^2 \delta t$ is very large. This allows the Poisson process $\delta
N(t)$  to be approximated by a Gaussian process:  
\beq \delta N(t) \simeq |\gamma|^2 \delta t +
|\gamma| \delta W(t), \eeq
where $\delta W(t)$ is a Gaussian random variable of zero mean and variance $\delta t$. Since
this time is infinitesimal as far as the system is concerned, we can substitute this
into equation (\ref{linqt1}) to obtain
 \beq \label{linqt2}
d\ket{\bar\psi(t)} = \left[ dW(t) e^{-i\Phi} c -  dt \left( iH + \half c\dg c 
\right) \right] \ket{\bar\psi(t)},
\eeq
where $\Phi = \arg \gamma$ and $dW(t)$ can be treated as an infinitesimal Wiener
increment \cite{Gar85} satisfying
\beq
{\rm E}[dW(t)]=0 \;,\;\; dW(t)^2 = dt.
\eeq
This equation has been previously derived by Goetsch and Graham, but in a
different way. Their method involved considering ideal measurements of one
quadrature of the radiated field, without involving photodetection or a local
oscillator. As there is no known physical mechanism for undertaking such an ideal
measurement, I believe that it is valuable to be able to derive the linear {\sc sse}
(\ref{linqt2}) as a limit of a physical detection scheme.

\section{Solving the Linear Quantum Trajectory}

It is now convenient to introduce a new
notation. Let the recorded photocurrent be defined
\beq
I(t)=\lim_{|\gamma|\to\infty} \frac{\delta N(t) - |\gamma|^2\delta
t}{|\gamma|\delta t} =  \frac{dW(t)}{dt},
\eeq
and let the complete record of the photocurrent from time 0 to just before time $t$
be denoted $\bfi$. Then as a reminder that $\ket{\bar\psi(t)}$ is conditioned on
the photocurrent, denote it by $\ket{\bar\psi^{\bf I}_t}$. Goetsch and Graham
\cite{GoeGra94b} have solved the linear {\sc SSE} (\ref{linqt2}) for the case of homodyne
detection, where the local oscillator phase $\Phi$ is held constant, and for the 
case of heterodyne detection where it varies rapidly: $\Phi(t) = \Phi_0 + \Delta
t$. This was done for some particularly simple internal Hamiltonians $H$. Here I
show that a solution may be found in a similar manner for any variation of $\Phi$
with time. I consider only the case $H=0$, which amounts to a freely
damped cavity. That is because I am interested in describing a single measurement
of some property of the initial state of the cavity, in which all of the light is
allowed to leak out and be detected.

Consider an initial coherent state $\ket{\psi_0} = \ket\alpha$ where
$a\ket{\alpha}=\alpha\ket{\alpha}$, where $a=c$ is the annihilation operator for
the damped mode, obeying $[a,a\dg]=1$. Now coherent states are unique with respect
to the damping master equation (\ref{241me2}) with $H=0$, in that they are the
only states which remain pure under that evolution equation. Specifically,
they have the solution $\ket{\psi_t}=\ket{\alpha e^{-t/2}}$. This means that the
solution to the linear {\sc SSE} (\ref{linqt1}) must be of the form
\beq
\ket{\bar\psi_t^{\bf I}} = \bar\psi_t^{\bf I} \ket{\alpha e^{-t/2}}.
\eeq
Now using the \ito calculus, equation (\ref{linqt2}) can also be written as
\beq
\ket{\bar\psi_{t+dt}^{\bf I}} = \exp(-\half a\dg a dt) \exp\bl dW(t)
e^{-i\Phi(t)} a - \half e^{-2i\Phi(t)} a^2 dt \br \ket{\bar\psi_t^{\bf I}}.
\eeq
Now using the fact that
\beq
\exp(-\half a\dg a t) \ket{\alpha} = \exp\bl -\half |\alpha|^2 (1-e^{-t}) \br
\ket{\alpha e^{-t}} ,
\eeq
one has
\beq
\bar\psi_{t+dt}^{\bf I} = \exp\bl -\half |\alpha|^2 e^{-t} dt - \half e^{-2i\Phi(t)}
\alpha^2 e^{-t}dt + e^{-i\Phi(t)} \alpha e^{-t/2} I(t) dt \br \bar\psi_t^{\bf I},
\eeq
where it is to be remembered that $dW(t)=I(t) dt$. The total solution satisfying
$\bar\psi_0 = 1$ is thus 
\beq
\bar\psi_t^{\bf I} = \exp \bl -\half |\alpha|^2 (1-e^{-t}) - \half \alpha^2 \int_0^t 
e^{-2i\Phi(s)}e^{-s}ds + \alpha \int_0^t e^{-i\Phi(s)} I(s) ds \br.
\eeq

Now it follows from this that the solution for the initial coherent state
$\ket{\psi_0}=\ket\alpha$ can also be written 
\beq \label{soln2}
\ket{\bar\psi_t^{\bf I}} = \exp(-\half a\dg a t) 
\exp \bl \half S^*(t) a^2  + R^*(t) a \br \ket{\psi_0},
\eeq
where
\beq
R(t) = \int_0^t e^{i\Phi(s)}e^{-s/2} I(s) ds \;;\;\;
S(t) = -\int_0^t e^{2i\Phi(s)}e^{-s} ds. \label{defRS}
\eeq
That is to say, the solution at time $t$ depends not on the full photocurrent
record $\bfi$ (which is a continuous infinity of real numbers), but
rather on that record only through the two complex integrals
(\ref{defRS}) \footnote{It might be thought that $S(t)$ does not even
depend on the record $\bfi$, but it might if the local oscillator phase $\Phi(t)$
is adjusted according to $\bfi$, so as to make an adaptive measurement as will be
considered below.}. Furthermore, because  equation (\ref{linqt1}) is a {\em linear}
equation by construction, the solution (\ref{soln2}) is true for arbitrary initial
conditions, not just for an initially coherent state. Thus any state conditioned on
photodetection in the large local oscillator limit depends on the photocurrent
only by the two integrals $R(t)$ and $S(t)$, regardless of how the local
oscillator phase varies. 

The measurement is complete in the limit $t\to\infty$, when there is no light in the
cavity and hence no information left to be obtained. In that limit,  $
\lim_{t\to \infty} \exp(-a\dg a t) = \ket{0}\bra{0}, $
so that the final conditioned state $\ket{0}$ is obviously of no interest. What is
of interest is the probability for having obtained the photocurrent record $\bfi$.
As we have seen above, the only relevant aspects of this record are $R(\infty)$,
which I will denote $A$, and $S(\infty)$, which I will denote $B$. From
Sec.~3, the probability for obtaining the results $A$ and $B$ is 
\beq
P(A,B) = P_0(A,B) \langle \bar\psi_\infty^{A,B} \ket{\bar\psi_\infty^{A,B}},
\eeq
where $\ket{\bar\psi_\infty^{A,B}}$ is the unnormalized conditioned state of
equation (\ref{soln2}) and $P_0(A,B)$ is the ostensible distribution for $A$ and $B$,
given by \beq \label{P0AB} P_0(A,B) = \int d{\bf I}_{[0,\infty)} P_0\bl {\bf
I}_{[0,\infty)}\br \delta^{(2)} \bl A-R(\infty)\br \delta^{(2)} \bl B -  S(\infty) \br.
\eeq
Here $R(\infty)$ and $S(\infty)$ are the functionals of ${\bf I}_{[0,\infty)}$
defined above, and $ P_0\bl {\bf I}_{[0,\infty)}\br$ is the ostensible
distribution for ${\bf I}_{[0,\infty)}$, which as seen above, is simply that of
Gaussian white noise.

From equation (\ref{soln2})  we
can rewrite the probability for obtaining results $A,B$ as
\beq
P(A,B) = {\rm Tr}[ F(A,B)\rho_0 ],
\eeq
where $\rho_0$ is the initial system state (which is allowed to be mixed), and
$F(A,B)$ is the effect for the results $A,B$ defined by
\beq \label{horesult}
F(A,B) = P_0(A,B) \bar\Omega\dg(A,B) \bar\Omega(A,B),
\eeq
where the (unnormalized) measurement operators are
\beq \label{genres}
\bar\Omega(A,B) = \ket{0}\bra{0} \exp \bl \half B^* a^2  + A^* a \br.
\eeq
This is a completely general result
describing the sorts of measurements which can be made on an optical mode by
photodetection in the large local oscillator limit. From it can be derived the
standard results for homodyne and heterodyne detection (as shown below),
and also some more interesting results for adaptive measurements (as I will discuss
in Sec.~6). At finite times one has less than maximal information about the
system and as a result the effect cannot be factorized in terms of measurement
operators. This is a minor generalization of the theory of operations and effects as
stated in Sec.~2 \cite{Gar91}; the same condition (\ref{coef}) for the effects applies. Explicitly,
equation (\ref{soln2}) shows that the {\sc povm} for the measurement up to time $t$ is 
\beq \label{foradapt}
F_t\bl R(t),S(t)\br = P_0 \bl R(t),S(t) \br  \exp \bl \half S(t) {a\dg}^2  + R(t)
a\dg \br \exp(-a\dg a t) \exp \bl \half S^*(t) a^2  + R^*(t) a \br,
\eeq
where $P_0\bl R(t),S(t) \br$ is defined analogously to (\ref{P0AB}).

\subsection{Homodyne Detection}

For homodyne detection of a given quadrature of the system, one desires the local
oscillator phase $\Phi$ to be constant. This means that $B$ is simply given by
$-e^{2i\Phi} \int_0^\infty e^{-t} dt = -e^{2i\Phi}$.
Thus it can be ignored as a measurement result, and we need consider only
\beq
A = e^{i\Phi} \int_0^\infty e^{-t/2} dW(t).
\eeq
Since $dW(t)$ is ostensibly white noise, it follows from equation (\ref{defRS}) that we
can write $A=e^{i\Phi}X$ where the ostensible distribution for $X$ is 
\beq
P_0(X)dX = \frac{1}{\sqrt{2\pi}} \exp(-X^2/2) dX.
\eeq
Thus the effect for such a measurement is
\beq \label{effhom}
F(X) = \frac{1}{\sqrt{2\pi}} \exp(-X^2/2)  \bar{E}(X) \ket{0}\bra{0}
\bar{E}\dg(X),
\eeq
where $\bar{E}(X)$ is an operator defined by
\beq
\bar{E}(X) = 
\exp\bl -\half e^{2i\Phi} a\dg{}^2 + X e^{i\Phi}a\dg \br 
\eeq

Now using some simple boson operator algebra \cite{Lou73}, it
is easy to show that  
\beq
(a e^{-i\Phi} + a\dg e^{i\Phi}) \bar{E}(X) \ket{0} = 
\bar{E}(X)  (a e^{-i\Phi} + X)\ket{0} = X \bar{E}(X) \ket{0}.
\eeq
That is to say, $\bar{E}(X)\ket{0}$ is an unnormalized eigenstate of 
of $a e^{-i\Phi} + a\dg e^{i\Phi}$ with eigenvalue $X$. The other factor in
equation (\ref{effhom}) supplies the normalization, so we can write 
\beq
F(X) = \ket{X}\bra{X},
\eeq
where $\ket{X}$ is an eigenstate of $a e^{-i\Phi} + a\dg e^{i\Phi}$ satisfying
the delta-function normalization $\langle X \ket{X'} = \delta(X-X')$. This is as
has been known for a long time; that homodyne detection can measure a quadrature
of the field arbitrarily accurately (see reference \cite{Wis95qo1} for a ``traditional''
derivation using quantum Langevin equations). Specific cases of this result have been
derived using the theory of quantum trajectories (see for example reference \cite{CarKocTia94}).
However, as far as I know, this is the first time it has been derived as a completely general
result using this theory.

\subsection{Heterodyne Detection}

Under heterodyne detection, the local oscillator phase $\Phi(t)$ varies rapidly
at a rate $\Delta \gg 1$. This means that $B$ averages over time to zero. The
remaining measurement result $A$ is defined by
\beq
A = \int_0^\infty e^{i\Phi(0)+i\Delta t-t/2} dW(t).
\eeq
It is easy to verify that in the limit $\Delta \gg 1$, this has the ostensible
distribution
\beq
P_0(A) d^2 A = \frac{1}{\pi} \exp(-|A|^2) d^2 A
\eeq
Thus the effect for a completed heterodyne measurement is
\beq
F(A) = \frac{1}{\pi}\exp(-|A|^2) \exp(a\dg A) \ket{0} \bra{0} \exp(a A^*).
\eeq
Using the techniques of reference \cite{Lou73}, it is again simple to show 
\beq
\exp(-|A|^2/2 + a\dg A) = \exp(a\dg A - a A^*)\exp(aA^*),
\eeq
and thus that
\beq
\exp(-|A|^2/2 + a\dg A)\ket{0} = 
\exp(a\dg A - a A^*)\ket{0} = \ket{A},
\eeq
where this last state is the normalized coherent state $a\ket{A} = A\ket{A}$. Thus
we have
\beq
F(A) = \frac{1}{\pi} \ket{A}\bra{A}.
\eeq
In this case the factor of $\pi^{-1}$ is necessary because the coherent states
are overcomplete. This result, that the distribution function for completed
heterodyne measurements is the $Q$ function $\pi^{-1} \bra{A}\rho \ket{A}$, has
also been known for some time (and again there is a succinct
derivation in reference \cite{Wis95qo1}). It has not previously been derived
using quantum trajectories as here.

\section{Adaptive Measurements} 

In the two examples considered above, homodyne and heterodyne measurement, the
quantum trajectory method for finding the finite time effect simply confirmed
already known results. A type of optical measurement for which the effects have
never before been calculated, and for which the quantum trajectory method is perhaps
the only tractable approach, is continuously adaptive measurements. These are
measurements in which the results so far (the photocurrent record $\bfi$ up to time
$t$) are used to alter the unitary matrix which determines the measurement operators
(\ref{urm}) at time $t$. Note that, as expressed in equation (\ref{aeie}), this does
not change the overall evolution of the system as given by the master equation
(\ref{241me2}), and so is distinct from feedback onto the system dynamics as explored
in Refs.~\cite{WisMil93b,WisMil94a,Wis94a}. In the infinite local oscillator amplitude
limit being considered here, there is only one parameter which determines
(\ref{urm}); that is the local oscillator phase $\Phi$. The feedback-control of this
phase is just what is required for a measurement which is of particular interest:
that of the optical phase of the system.

A standard measurement of the phase of an optical system consists of making a
completed heterodyne measurement yielding the result $A$ as above, and then
discarding the amplitude information in $A$. That is to say, the effect for such a
standard phase measurement is
\beq
F_{\rm std}(\phi) = \int |A|d|A| \frac{1}{\pi} \ket{A}\bra{A},
\eeq
where $A=|A|e^{i\phi}$. Because it discards information, this effect cannot be
factorized into measurement operators as in equation (\ref{hereisfac}). In the
language of reference \cite{Uff93}, it is an unsharp measurement of phase. By contrast,
an ideal measurement of phase \cite{Lon27,PSphase93,LeoVacBohPau95} has the
effect 
\begin{equation} \label{Fopt1}
 {F}_{\rm ideal}(\phi)=\frac{1}{2\pi} |{\phi}\rangle\langle{\phi}| \; , \;\;
|{\phi}\rangle = \sum_{n=0}^\infty e^{i n \phi} |{n}\rangle,
\end{equation}
defined in terms of the unnormalized phase eigenstates $\ket\phi$. Ideal measurements
are the optimal way to extract information which has been encoded in the phase of the
field \cite{HalFus91}.

Standard measurements of the phase are suboptimal because they measure both
the phase and amplitude quadratures of the field, but use only the former
information. It is intuitively obvious that one could make a better measurement 
(i.e. closer to ideal) by concentrating on measuring the phase quadrature.
However assuming that one knows the phase quadrature before making the
measurement is not in the spirit of a true phase measurement. Rather, one would
have to estimate the phase quadrature during the course of the measurement, from
the photocurrent so far. This is the essence of the adaptive measurement of
phase. Precisely what algorithm one used to estimate the phase quadrature would
depend on the knowledge one had about the initial state of the system.
Nevertheless, the analysis given above shows that whatever algorithm one used,
the estimated phase at time $t$, $\hat\varphi_t$ should depend on the
photocurrent $\bfi$ only through the two complex numbers $S(t)$ and $R(t)$. In
order to measure the phase quadrature, the local oscillator phase $\Phi(t)$ should
therefore be set equal to 
\beq \label{ctrl}
\Phi(t) = \hat\varphi_t[R(t),S(t)] + \pi/2.
\eeq
When the measurement is over (at $t=\infty$), one's estimate for the original phase of the
system is $\phi=\hat\varphi_\infty$. 

Exactly how $\hat\varphi_t$ should be chosen as a function of $R(t)$ and $S(t)$
is a complicated issue. As stated above, it depends on the knowledge the observer
has regarding the preparation of the system. This is evident in figure 1, which 
shows the  Wigner function of the effect
 $F_t\bl R(t),S(t)\br$ (\ref{foradapt}) for some typical
values of $R$ and $S$ which might occur in an adaptive phase measurement. The
position and shape of this positive operator in phase space ($q=a+a\dg$ and $p =
-ia+ia\dg$) indicates the inherent uncertainties in the results of the single
measurement up to time $t$ yielding results $R$ and $S$. As this figure shows, the
state $F_t\bl R(t),S(t)\br$ is a Gaussian state. As $t\to\infty$ it becomes a
minimum uncertainty squeezed state. For heterodyne detection this reduces to a
coherent state of complex amplitude $A$. For homodyne detection it is an
eigenstate of $p\cos\theta+q\sin\theta$, where $\theta$ is in this case
equal to the chosen local oscillator phase $\Phi$. In general $\theta=\arg(S)/2$
is a measurement result which changes over the course of the adaptive measurement.
The displacements $x$ and $y$ are defined in terms of the
measurement results $R$ and $S$ by  \beq x = \frac{2 \Re (R
e^{-i\theta})}{1-e^{-t}-|S|} \;;\;\;  y = \frac{2 \Im (R
e^{-i\theta})}{1-e^{-t}+|S|}. \label{defxy} \eeq  The variances in the $x$ and $y$
quadrature are \beq V_x = \frac{1-e^{-t}+|S|}{1-e^{-t}-|S|} \;;\;\;
 V_y = \frac{1-e^{-t}-|S|}{1-e^{-t}+|S|}. \label{defvxvy} \eeq 
 Evidently, the phase of this state is not unambiguously defined in general. One
reasonable estimate would be the phase of the centre of the Wigner function, which
would be  \beq
\hat\varphi_t = \theta_t + \arctan(y_t/x_t) = \arg \left[ R(1-e^{-t}) + S R^* \right].
\eeq
However, depending on one's initial knowledge of the state of the
system, a different estimate may be more appropriate. 

There is one example for which a different estimate is definitely more
appropriate, and for which there is no ambiguity about the estimated phase of the
system. This is when there is at most one photon in the system (and the observer
knows this).  First, I wish to present the results in this case for ideal and
standard phase measurements. Projecting the optimal mode effects into the subspace
spanned by $\{\ket{0},\ket{1}\}$ gives the ideal phase measurement effect
\begin{equation} {F}_{\rm ideal}(\phi) = \frac{1}{2\pi} |{\phi}\rangle\langle{\phi}|
\; , \;\;  |{\phi}\rangle =
|{0}\rangle + e^{i\phi}|{1}\rangle, \end{equation}
and the marginal heterodyne effect
\beq
F_{\rm std}(\phi) = \frac{\sqrt{\pi}}{2} 
{F}_{\rm ideal}(\phi) + \left(1-\frac{\sqrt{\pi}}{2}\right)\frac{{1}}{2\pi}. 
\label{std2}
\eeq
That is to say, the standard technique has an efficiency of $\sqrt{\pi}/2 \simeq
88\%$, in the sense that the same {\sc povm} would arise from a ideal measurement
which worked $88\%$ of the time, and which gave a random answer on the interval
$[0,2\pi)$ the other $22\%$ of the time. 

I have considered the adaptive phase measurement of a field with at most one photon
in detail in reference \cite{Wis95prl}. The lack of ambiguity in the estimated
phase quadrature $\hat\varphi_t$ is due to the fact that with at most one photon
in the field, the measurement result $S(t)$ is irrelevant to the
estimation of its state, which is obvious from equation (\ref{soln2}). The estimated
phase quadrature in this case is $\hat\varphi_t = \arg R(t)$. Substituting this
into equation (\ref{ctrl}) leads to an analytic solution for $P_0(A)$:
\begin{equation}
P_0(A)d^2A = \pi^{-1} \delta(|A|^2 - 1)d^2A.
\end{equation}
In other words, $A$ can contain only phase information, not amplitude information.
This is precisely what is desired for a phase measurement, and indeed the {\sc povm} for
this adaptive measurement is \begin{equation}
{F}_{\rm adapt}(\phi) = \int_0^\infty |A| d|A|
P_0(A) (1+\ket1\bra0 A) \ket{0}\bra{0} (1+\ket0\bra1 A^*) = {F}_{\rm ideal}(\phi),
\end{equation} Here I have again used the result (\ref{horesult}) for an optical
mode, projected into the lowest two levels $\{\ket{0},\ket{1}\}$. Thus for any
field with at most one photon,  a simple adaptive measurement scheme can produce an
optimal measurement of the phase, which is significantly better than the result
with no feedback (\ref{std2}).

\section{Conclusion}

In this paper I have investigated the reciprocal links between quantum
trajectories and quantum measurement theory. The first link is that one
interpretation of quantum trajectories is as a special case of quantum
measurement theory, that of continuous observation. Here each measurement
has a duration which is infinitesimal on the time scale of the non-trivial system
evolution, and is described in terms of operations and effects.
This leads naturally to discrete measurement results (photodetections). The
equations are actually easy to solve in the limit of a very large local
oscillator, in which case the discrete photodetections are replaced by a noisy
photocurrent. This is the limit which was considered in detail here.

The second link is that by solving the quantum trajectories over a finite time
one can obtain the operations and effects for measurements of finite duration. These
results would be difficult to obtain were it not for a formulation of quantum
trajectories as linear stochastic \sch equations due to Goetsch and Graham which I
have rederived here in a new way. For the case where the the system is not driven, but 
simply allowed
to  radiatively damp, the finite measurement time is naturally extended to infinity in order to
detect all of the radiated light. Then it is possible to rederive the standard results for
homodyne and heterodyne detection, and also to derive new results which would
be very difficult to derive by other means. These new results are those
pertaining to adaptive measurements, in which, for example, the photocurrent up to
time $t$ is used to control the local oscillator phase at time $t$. Here I
presented the results for one case which can be completely solved analytically: adaptive phase
measurements on a system with an upper photon number bound of one.

The topic of real-time adaptive measurements begins rather than ends with phase
measurements. Feedback-control of measurements can only improve the quality with
which properties are measurable compared with standard non-adaptive techniques.
For measuring quantities other than phase, one would not wish to work in the
infinite local oscillator amplitude limit. Rather, one would have a finite local
oscillator, in which case there would be two parameters to be controlled: its
amplitude and phase. This is obviously a more difficult problem, and it may be
that only a numerical solution using stochastic \sch equations would be possible.
Even using numerical techniques, it is not obvious what the general algorithm
would be for optimizing such measurements, nor even how optimization should be
defined. These are problems for future work, continuing the fruitful interaction
between formal quantum measurement theory and its principal realm of application
in quantum optics.

\section*{Acknowledgments}

I would like to acknowledge valuable discussions with S.M. Tan.
This project was supported by the New Zealand Foundation for Research,
Science and Technology.

\newpage

\begin{figure}
\psfig{figure=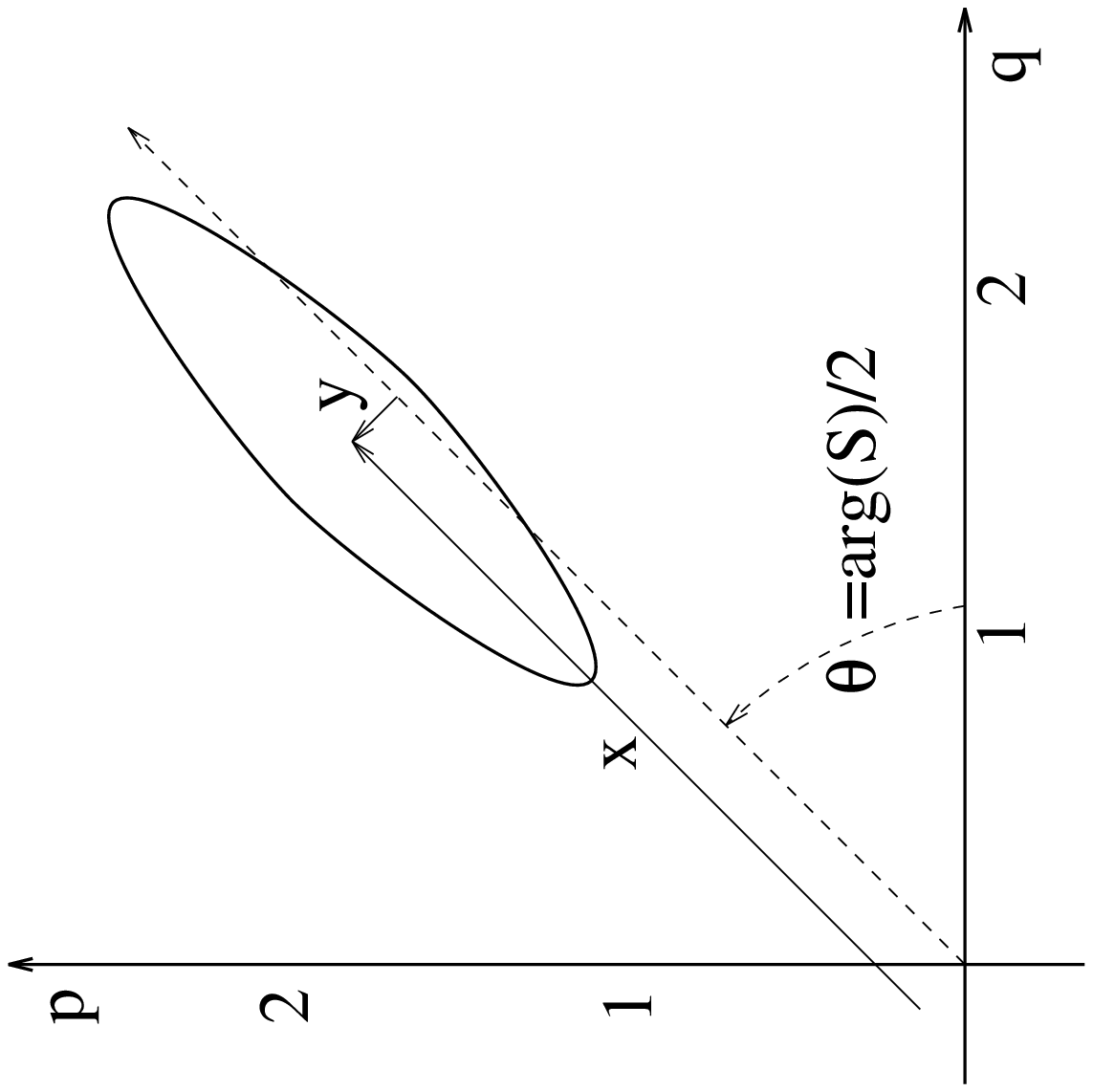,width=65mm}
\caption
{One-standard-deviation contour for the Wigner function of the effect
$F_t(R,S)$  (56)
for typical values of $R$ and $S$ which would occur in
the adaptive measurement of the phase of a state with a moderate coherent
amplitude. Here $q=a+a\dg$ and $p = -ia+ia\dg$, $x$ and $y$ are defined by
equation (72),
and the variances in these quadratures by equation (73).}
\end{figure}


\begin{thebibliography}{99}


\bibitem{DalCasMol92}
J. Dalibard, Y. Castin and K. M\o lmer,
Phys. Rev. Lett. {\bf 68}, 580 (1992).

\bibitem{Car93b}
H.J. Carmichael,
{\em An Open Systems Approach to Quantum Optics}
(Springer-Verlag, Berlin, 1993).

\bibitem{Boh58}
N. Bohr,
``Quantum physics and philosophy --- causality and complementarity''
in {\em Essays on Atomic Physics and Human Knowledge}
(Wiley, New York, 1958).

\bibitem{Hei59}
W. Heisenberg,
{\em Physics and Philosophy}
(Allen \& Unwin, London, 1958).

\bibitem{MolCasDal93}
K. M\o lmer, Y. Castin, and J. Dalibard,
J. Opt. Soc. Am. B {\bf 10}, 524 (1993).

\bibitem{DumZolRit92}
R. Dum, P. Zoller and H. Ritsch,
Phys. Rev. A {\bf 45}, 4879 (1992).

\bibitem{MarDumTaiZol93}
P. Marte, R. Dum, R. Ta\"{\i}eb and P. Zoller,
Phys. Rev. A {\bf 47}, 1378 (1993).

\bibitem{Mar93}
P. Marte {\em et al}
Phys. Rev. Lett. {\bf 71}, 1335 (1993).

\bibitem{HolMarMarZol95}
M. Holland, S. Marksteiner, P. Marte and P. Zoller,
submitted to Phys. Rev. Lett. (1995).

\bibitem{AlsCar91}
P. Alsing and H. J. Carmichael,
Quantum Optics {\bf 3}, 13 (1991).

\bibitem{TiaCar92} 
L. Tian and H.J. Carmichael,
Phys. Rev. A {\bf 46}, R6801 (1992).

\bibitem{Car93a}
H.J. Carmichael,
Phys. Rev. Lett. {\bf 70}, 2273 (1993).

\bibitem{CarKocTia94}
H.J. Carmichael, P. Kochan, and L. Tian,
in {\em Proceedings of the International Symposium on Coherent 
States: Past, Present, and Future}
(World Scientific, Singapore, 1994).

\bibitem{KocCarMorRai95}
P. Kochan, H.J. Carmichael, P.R. Morrow and M.G. Raizen,
Phys. Rev. Lett. {\bf 75}, 45 (1995).

\bibitem{GarParZol92}
C.W. Gardiner, A.S. Parkins, and P. Zoller,
Phys. Rev. A {\bf 46}, 4363 (1992).

\bibitem{SriDav81}
M.D. Srinivas and E.B. Davies,
Opt. Acta {\bf 28}, 981 (1981).

\bibitem{CarSinVyaRic89}
H.J. Carmichael, S. Singh, R. Vyas, and P.R. Rice,
Phys. Rev. A {\bf 39}, 1200 (1989).

\bibitem{Coo88}
R.J. Cook, 
Phys. Scripta {\bf T21}, 49 (1988).

\bibitem{GagMil93} 
M.J. Gagen and G.J. Milburn,
Phys. Rev. A {\bf 47}, 1467 (1993).

\bibitem{WisMil93c} 
H.M. Wiseman and G.J. Milburn,
Phys. Rev. A {\bf 47}, 1652  (1993).

\bibitem{GoeGra94b}
P. Goetsch and R. Graham,
Phys. Rev. A {\bf 50}, 5242 (1994).

\bibitem{GoeGraHaa95a}
P. Goetsch, R. Graham and F. Haake,
Phys. Rev. A {\bf 51}, 136 (1995).

\bibitem{GarKni94}
B.M. Garraway and P.L. Knight,
Phys. Rev. A {\bf 50}, 2548 (1994).

\bibitem{QuaColWal95}
R. Quadt, M.J. Collett, and D.F. Walls,
Phys. Rev. Lett. {\bf 74}, 351 (1995).

\bibitem{WisMil93b} 
H.M. Wiseman and G.J. Milburn,
Phys. Rev. Lett. {\bf 70}, 548 (1993).

\bibitem{WisMil94a} 
H.M. Wiseman and G.J. Milburn,
Phys. Rev. A {\bf 49}, 1350 (1994).

\bibitem{Wis94a} 
H.M. Wiseman,
Phys. Rev. A {\bf 49}, 2133 (1994);
Errata {\em ibid.}, {\bf 49} 5159 (1994) and {\em ibid.} {\bf 50}, 4428 (1994).

\bibitem{WisMil94b} 
H.M. Wiseman and G.J. Milburn,
Phys. Rev. A {\bf 49}, 4110 (1994).

\bibitem{LieMil95}
A. Liebman and G.J. Milburn,
Phys. Rev. A {\bf 51}, 736 (1995).

\bibitem{Wis95a} 
H.M. Wiseman,
Phys. Rev. A {\bf 51}, 2459 (1995).

\bibitem{Wis95b}
H.M. Wiseman,
Modern Physics Lett. B {\bf 9}, 629 (1995).

\bibitem{Gis89}
N. Gisin, 
Helv. Phys. Acta {\bf 62}, 363 (1989).

\bibitem{Dio88}
L. Diosi, 
J. Phys. A {\bf 21}, 2885 (1988).

\bibitem{Pea89}
P. Pearle,
Phys. Rev. A {\bf 39}, 2277 (1989).

\bibitem{GRW}
G.C. Ghirardi, A. Rimini, and T. Weber, 
Phys Rev. A {\bf 34}, 470 (1986).

\bibitem{Roh87}
F. Rohrlich,
Found. Phys. {\bf 17}, 1205 (1987).

\bibitem{TeiMah92}
W.G. Teich and G. Mahler,
Phys. Rev. A {\bf 45}, 3300 (1992).

\bibitem{GisPer92a}
N. Gisin and I. Percival,
Phys. Lett. A {\bf 167}, 315 (1992).

\bibitem{GisPer92b}
N. Gisin and I. Percival,
J. Phys. A {\bf 25}, 5677 (1992).

\bibitem{GisPer93}
N. Gisin and I. Percival,
J. Phys. A {\bf 26}, 2233 (1993); 
{\em ibid.}, 2245 (1993).

\bibitem{Gis93}
N.~Gisin, P.L.~Knight, I.C.~Percival, R.C.~Thompson, and 
D.C.~Wilson,
J. Mod. Optics {\bf 40}, 1663 (1993).

\bibitem{BrePet95a}
H-P. Breuer and F. Petruccione,
Phys. Rev. Lett. {\bf 74}, 3788 (1995).

\bibitem{BrePet95b}
H-P. Breuer and F. Petruccione,
Phys. Rev. E {\bf 52}, 428 (1995).

\bibitem{Gar91}
C.W. Gardiner,
{\em Quantum Noise}
(Springer-Verlag, Berlin, 1991).

\bibitem{Pen90}
R. Penrose,
{\em The Emperor's New Mind}
(Vintage, London, 1990)

\bibitem{Per94}
I. Percival,
Proc. Roy. Soc. A {\bf 447}, 189 (1994).

\bibitem{Von32}
J. von Neumann, {\em Mathematical Foundations of Quantum Mechanics}
(Springer, Berlin, 1932);
English translation (Princeton University Press, Princeton, 1955).

\bibitem{Zur82}
W.H. Zurek,
Phys. Rev. D {\bf 26}, 1862 (1982).

\bibitem{MisSud77}
B. Misra and E.C.G. Sudarshan,
J. Math. Phys. {\bf 18}, 756 (1977).

\bibitem{Dav76}
E.B. Davies,
{\em Quantum Theory of Open Systems}
(Academic Press, London, 1976).

\bibitem{Gar85}
C.W. Gardiner,
\newblock {\em Handbook of Stochastic Methods}
\newblock (Springer-Verlag, Berlin, 1985).

\bibitem{Lin76}
G. Lindblad,
Commun. math. Phys. {\bf 48}, 199 (1976).

\bibitem{CoxIsh80}
D.R. Cox and V. Isham,
{\em Point Processes}
(Chapman and Hall, London, 1980).

\bibitem{Boh13}
N. Bohr, 
Phil. Mag. {\bf 26}, 1 (1913).

\bibitem{Lou73}
W.H. Louisell,
{\em Quantum Statistical Properties of Radiation}
(Wiley, New York, 1973).

\bibitem{Wis95qo1}
H.M. Wiseman,
Quantum and Semiclassical Optics (JEOS B) {\bf 7}, 569 (1995).

\bibitem{Uff93}
J. Uffink,
Int. J. Theor. Phys. {\bf 33}, 199 (1994).

\bibitem{Lon27}
F. London, 
Z. Phys. {\bf 40}, 193 (1927).

\bibitem{PSphase93}
Physica Scripta {\bf T48}, 13 (1993)
{\em Quantum Phase and Phase Dependent Measurements}
edited by W.P. Schleich and S.M. Barnett.

\bibitem{LeoVacBohPau95}
U. Leonhardt, J.A. Vaccora, B. B{\"o}hmer, and H. Paul,
Phys. Rev. A {\bf 51}, 84 (1995).

\bibitem{HalFus91}
M.J. Hall and I.G. Fuss,
Quantum Opt. {\bf 3}, 147 (1991).

\bibitem{Wis95prl}
H.M. Wiseman,
Phys. Rev. Lett. {\bf 75}, 4587 (1995).


\end{thebibliography}
\end{document}